\documentclass[fleqn]{article}
\usepackage[utf8]{inputenc}
\usepackage[english]{babel}

\usepackage{bbm}
\usepackage{amsmath, amsthm, amssymb, amsfonts}
\usepackage{mathrsfs}
\usepackage{rotating}
\usepackage{graphicx}
\usepackage[colorlinks]{hyperref}
\usepackage{listings }
\usepackage{fancyvrb}

\newtheorem{theorem}{Theorem}

\title{Calculating permutation entropy without permutations}
\author{Alexander K.Vidybida\thanks{\tt vidybida@bitp.kiev.ua, http://vidybida.kiev.ua}\\ {\small\em Bogolyubov Institute for Theoretical Physics} \\ 
{\small\em 14-b Metrolohichna str. Kyiv, 03143, Ukraine}}

\begin{document}
\maketitle

%\hrule
\begin{abstract} 
A method for analyzing sequential data sets, similar to the permutation entropy one,
is discussed. The characteristic features of this method are as follows:
it preserves information about equal values, if any, in the embedding vectors; 
it is exempt of combinatorics;
it delivers the same entropy value as does the permutation method,
provided the embedding vectors do not have equal components.
In the latter case this method can be used instead of the permutation one.
If embedding vectors have equal components this method could be more
precise in discriminating between similar data sets.
\medskip

\end{abstract}

%\hrule
\bigskip

\section{Introduction}
Due to technical 
progress in the areas of sensors and storage devices a huge amount of raw data about time course of different processes, such as ECG, EEG, climate data recordings, stock market data have become available. 
These data are redundant. The data processing and classification, aimed at 
extracting meaningful for nonspecialist characteristics, is based on reducing the excess of redundancy.
As a result, a new data is obtained, small in size and digestible by a human being.
Examples of those reduced data for time series can be mean value, variance, 
Liapunov exponents, correlation dimension, attractor dimension and others.

%\paragraph{Outline}
A remarkable method suitable for reducing the excess of redundancy in time series
 has been proposed by Ch.Bandt and B.Pompe in
\cite{Bandt2002}, known as permutation entropy. This method is simple and transparent, is robust with respect to monotonic distortions of the raw data, and is suitable for estimating the dynamical complexity
of the underlying dynamical process. Many interesting results, e.g. 
\cite{Porta2001,Zanin2012,Bariviera2015,Tylov2018}, have been obtained with
 straightforward application of the permutation entropy methodology in its initial form, as it is 
described in \cite{Bandt2002}.
Nevertheless, this method is subjected to a critique for not taking into account absolute values of
the raw data and for not treating properly a possibility of having equal values in the embedding vector (ties),
\cite{Zunino2017,Cuesta-Frau2018}. In this connection, it should be taken into account that any
 redundancy
reduction method leaves out some type of information, which may be useless for one process/task and may carry
useful information for another one. In the latter case, the bare idea of \cite{Bandt2002}
about how to treat equal values can/should be 
modified in order to meet a purpose of concrete situation. Examples of such a modification can be found
in \cite{Azami2016,Chen2019} for taking into account absolute values, or in \cite{Bian2012,Haruna2013} for treating equal values. Interesting modification of the permutation entropy method has been proposed in
 \cite{Berger2017} for 3‑tuple EEG data. 
 
In the standard permutation entropy methodology, it is preferable that embedding vectors
 have all their components different. Otherwise, they cannot be plainly symbolized by a permutation
 without using additional rules, which actually treat equal values as not being such.
 Situation with equal values in the embedding vector may arise for high embedding dimension,
 for crude quantization of measured data, for very long data sequences 
 and when observed dynamical system has intrinsically
 only a small number of possible outputs. 
 
 This note is aimed at discussing a slightly different symbolization technique of embedding vectors,
 which does not refer to combinatorics, and which is capable of preserving information about equal values
 in embedding vectors. Instead of permutation, an embedding vector is emblematized with a single integer number of base $D$, where $D$ is the embedding dimension. In the case of no ties (no equal components in the embedding vectors) the technique is equivalent to the standard permutation entropy methodology.
 In the opposite case, it may discriminate between similar data sets better than the permutation
 entropy method does.

\section{Permutation entropy}\label{P}

Consider a finite sequence 
\begin{equation}\label{raw}
\mathsf{X} = (x_0,x_1,\dots, x_{N-1}),\quad x_i\in \mathbb{R}^1, \quad i=0,1,2,\dots N-1,
\end{equation}
of measurements. By choosing the embedding dimension $D<N$ the data
 (\ref{raw}) can be embedded into a $D$-dimensional space by picking out consecutive
 $D$-tuples from $\mathsf{X}$. As a result, a set of $D$-dimensional 
 embedding vectors is obtained:
 \begin{equation}\label{Draw}
\mathbf{V} = \{V_0,V_1,\dots, V_{N-D}\},\quad V_i\in \mathbb{R}^D, \quad i=0,1,2,\dots N-D,
\end{equation}
where each vector has the following form:
\begin{align}\nonumber
V_0 =& \,(x_0,x_1,x_2\dots, x_{D-1}),\quad\dots\quad, 
\\\label{Dvec}
V_i=& \,(x_i,x_{i+1},x_{i+2},\dots, x_{i+D-1}),\quad\dots\quad ,
\\\nonumber
V_{N-D}=& \,(x_{N-D},x_{N+1-D},\dots, x_{N-1}).
\end{align}
An additional parameter of the embedding procedure is delay $\tau =1,2,\dots$. In the above definition,
we put $\tau=1$ for simplicity. With $\tau\ne1$ one would have 
$V_i= (x_i,x_{i+\tau},x_{i+2\tau},\dots, x_{i+(D-1)\tau})$ instead of (\ref{Dvec}).

The data represented in (\ref{Draw}) and/or (\ref{Dvec}) is even more redundant than that
represented in (\ref{raw}) since, for $D\ll N$, most data values from (\ref{raw}) are represented in (\ref{Dvec})
$D$ times. In the permutation entropy technique \cite{Bandt2002}, each embedding vector from
(\ref{Draw}) and/or (\ref{Dvec}) is replaced with a permutation $\pi$ of $D$ integers \{0,1,2,\dots,$D-1$\},
which is defined by the order pattern of values composing the vector. For any embedding vector
$V=(x_0,x_1,\dots,x_{D-1})$ the permutation $\pi$, which symbolizes it,
 is calculated as follows. Arrange all components
of $V$ either in the descending, \cite{Keller2014}:
\begin{equation}\label{desc}
V=(x_0,x_1,\dots,x_{D-1}) \rightarrow V_\pi=(x_{r_0},x_{r_1},\dots,x_{r_{D-1}}),\,
x_{r_0}>x_{r_1}>\dots>x_{r_{D-1}},
\end{equation}
or in the ascending, \cite{Haruna2013,Gutjahr2020}:
\begin{equation}\label{asc}
V=(x_0,x_1,\dots,x_{D-1}) \rightarrow V_\pi=(x_{r_0},x_{r_1},\dots,x_{r_{D-1}}),\,
x_{r_0}<x_{r_1}<\dots<x_{r_{D-1}}
\end{equation}
order\footnote{Actually, in (\ref{desc}), (\ref{asc}) equal values (ties) are as well admitted.
Here, we exclude such a possibility for the sake of clarity. The equal values are discussed in 
the next section.} keeping their subscripts unchanged. 
The permutation $\pi$ which corresponds to $V$ is obtained as the row of the subscripts in the 
rearranged vector $V_\pi$ from either (\ref{desc}), or (\ref{asc}):
\begin{equation}\label{pi}
\pi\equiv\pi(V)=(r_0,r_1,\dots,r_{D-1}).
\end{equation}
From the set of embedding vectors $\mathbf{V}$, calculate a new set $\Pi$ of order
patterns by replacing each vector in (\ref{Draw}) by the corresponding permutation:
\begin{equation}\nonumber %\label{Pi}
\Pi=\,\{\pi_0,\pi_1,\dots,\pi_{N-D}\}.
\end{equation}
Now, empirical probability of each permutation, $p(\pi_i)$, can be obtained by 
dividing the number of occurrences of $\pi_i$ in the $\Pi$ by the total number of elements
in the $\Pi$. The permutation entropy of $\mathbf{V}$ is the Shannon entropy of the probability
distribution $p(\pi_i)$:
\begin{equation}\nonumber %\label{PE}
H(\mathbf{V})\equiv H(\Pi)=-\sum\limits_{i=0}^{K-1}p(\pi_i)\log(p(\pi_i)),
\end{equation}
where $K$ is the number of different permutations in the $\Pi$.

\subsection{Treatment of equal values}\label{equal0}

\begin{figure}[t]
\includegraphics[width=0.3\textwidth]{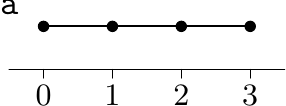}\hfill
\includegraphics[width=0.3\textwidth]{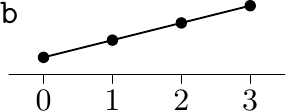}\hfill
\includegraphics[width=0.3\textwidth]{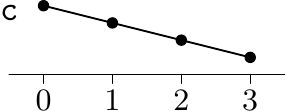}
\caption{\label{asdes}In the standard permutation entropy symbolization, 
a sequence of same values ({\tt a}) is
equivalent either to ascending ({\tt b}), or descending ({\tt c}) sequence, if either (\ref{desc}),
or (\ref{asc}) convention is used. }
\end{figure}

Equal values in an embedding vector are, to an extent, inconvenient.
Indeed, if $x_{r}=x_{s}$ for some $0\le r,s< D$ in a vector $V=(x_0,x_1,\dots,x_{D-1})$, 
then $r$ and $s$ should be placed side by side in the permutation (\ref{pi}), 
but which one should go first?
Due to sameness of values it is impossible to uniquely determine a corresponding permutation
without introducing additional rules.
In some cases the possibility of equal values can be ignored due to their low probability.
This is reasonable when the embedding dimension is low, and/or a chaotic process data are
recorded with high precision, see \cite{Bandt2002,Bandt2005,Aziz2005}.
If equal values are inevitable, the following rule is applied\footnote{In some cases, e.g. \cite{Bian2012,Haruna2013}, the opposite inequality sign is used here.}
\begin{equation}\label{rule}
\text{\bf if }x_s=x_r \text{ \bf and } s>r \text{ \bf than } s \text{ goes first.}
\end{equation}
The rule (\ref{rule}) has different meaning depending of whether (\ref{desc}) or (\ref{asc})
convention is used. Namely, in the case of (\ref{desc}), an embedding vector with all
components equal will be equivalent to a vector with monotonically ascending components.
If (\ref{asc}) is adopted, then that same vector will be equivalent to a vector with
 monotonically descending components, see Fig. \ref{asdes}.
Without knowing a real system, it is not clear which case is better and whether it is
good or bad to label a sequence of same values as being decreasing or increasing.
Actually, the permutation symbolization technique aims at reducing redundancy.
Discrimination between constant and either increasing, or decreasing
sequences of data may appear to be excessive in some cases. 
On the other hand, when a system
generating data has a few possible outputs, or the data was subjected to a crude 
quantization, or embedding dimension is large,
 it may happen to be useful if presence of
 equal values in the embedding vector results in order pattern preserving this fact.
 One possible approach to do this is discussed in the next section.

\section{Arithmetic entropy}
\subsection{Symbolization}\label{Sym}
The following symbolization is aimed to keep information about equal values
in embedding vectors. Having a vector $V=(x_0,x_1,\dots,x_{D-1})$ construct a
sequence of integers $\alpha$:
\begin{equation}\label{aldef}
V=(x_0,x_1,\dots,x_{D-1}) \rightarrow \alpha=\alpha(V)=(a_0,a_1,\dots,a_{D-1})
\end{equation}
 by the following
rule. Find the smallest component, $c_0$, in the $V$. If $c_0$ is found at places 
$r_1,r_2,\dots$, put number 0 at those places in the $\alpha(V)$. Find the next smallest
component $c_1$, $c_1>c_0$ in the $V$. If $c_1$ is found at places 
$s_1,s_2,\dots$, put number 1 at those places in the $\alpha$. Proceed this way until
 components of $V$ are exhausted. At this stage, all $D$ components of $\alpha$ will be
determined. The $\alpha$ obtained this way is used as a symbol of embedding vector $V$.

For example, consider $V=(7, 15, 7, 25, 15)$. The corresponding symbol, or the order pattern
is $\alpha=(0,1,0,2,1)$. Here, information about equal values and their positions is
preserved.

If $V$ has no equal components, it can be proven (see Appendix \ref{pi-1}) that $\alpha=\pi^{-1}$. 
This means that $\alpha$ is the 
inverse permutation of the one obtained for $V$ if convention (\ref{asc}) is used.
Since correspondence between permutations and their inverse is one-to-one, it does not matter
which one, $\pi$ or $\alpha$, is used for calculating entropy. This further means that for a data set
and embedding method, which does not deliver equal values
in the embedding vectors, symbolization used here is equivalent 
to the permutation one\footnote{It seems, that in paper \cite{Tylov2018} symbolization method described
 here is used. But, as it may be concluded
from \cite[Eq. (6)]{Tylov2018}, the issue of equal values is not addressed. Similar
approach is used in \cite{Kulp2014,Berger2017}, again without considering equal values.} while calculating 
entropy.

\subsection{Arithmetization}\label{Ari}

Expect that embedding vector $V$ in (\ref{aldef}) has exactly $d$ unique components,
where $d\le D$. In this case, corresponding symbol $\alpha(V)$ will be a sequence 
of $D$ numbers chosen from the set $\mathbbmss{d}=\{0,1,\dots,d-1\}$ in such a way that 
not any element from $\mathbbmss{d}$ is missed. The latter can be formulated as the following condition:
\begin{equation}\label{cond}
\bigwedge_{b\in \mathbbmss{d}}\,\,b\in \alpha(V).
\end{equation}

The sequence $\alpha(V)$ can be considered as a single integer $A(V)$, 
in a base-$D$ positional numeral system\footnote{For a single embedding vector, $d$ might be chosen
as radix instead of $D$. But $d$ may be different for different vectors. And a same integer may have different representation for different bases with (\ref{cond}) satisfied. E.g. 0112$_3$ = 1110$_2$.}, with digits 
$a_{D-1}a_{D-2}\dots a_0$:
\begin{equation}\label{A}
A(V)\equiv A =a_0 + a_1\,D + a_2\,D^2 +\dots + a_{D-1}D^{D-1}.
\end{equation}
It is clear that there is one-to-one correspondence between order patterns $\alpha$ and 
integers obtained as shown in (\ref{A}).
Therefore, a set of order patterns, constructed as described in Sec. \ref{Sym},
can be replaced with a set $\mathcal{A}$ of integers obtained as shown in (\ref{A}):
\begin{equation}\label{AA}
\mathcal{A}=\{A_0, A_1,\dots, A_{N-D}\}, \text{ where } A_i\equiv A(V_i).
\end{equation}
The empirical probabilities $p(A_i)$ to find an integer $A_i$ among those in $\mathcal{A}$ can be calculated
as usual, and we have for the arithmetic entropy:
\begin{equation}\nonumber %\label{AE}
H_a(\mathbf{V})\equiv H_a(\mathcal{A})=-\sum\limits_{i=0}^{L-1}p(A_i)\log(p(A_i)),
\end{equation}
where $L$ is the number of different integers in the $\mathcal{A}$.

For a data sequence and embedding method which does not deliver equal values in the embedding vectors,
all $d_i=D$ and the integers $A_i$ will represent corresponding permutation order patterns
unambiguously. In this case, $A_{min}\le A_i \le A_{max}$, where $A_{min}$ corresponds to
pattern $\alpha_{min}=(D-1,D-2,\dots,1,0)$:
$$
A_{min}=D-1 + (D-2)\,D + (D-3)\,D^2 +\dots+D^{D-2},
$$
and $A_{max}$ corresponds to pattern $\alpha_{max}=(0,1,\dots,D-1)$:
$$
A_{max}=D + 2\,D^2 +3\,D^3+\dots+(D-1)\,D^{D-1}.
$$
In this case, only $D!$ integers will be used from $[A_{min}; A_{max}]$ due to condition (\ref{cond}).

\subsection{How many new possible order patterns are got?}\label{got}

If it is decided to treat order patterns generated from embedding $D$-vectors with some 
components equal as not equivalent to those from vectors with all components different,
then the number of all possible patterns will be greater than $D!$. Here we attempt to
estimate how many new patterns can be obtained.

\begin{figure}
\begin{center}
\includegraphics[width=0.6\textwidth]{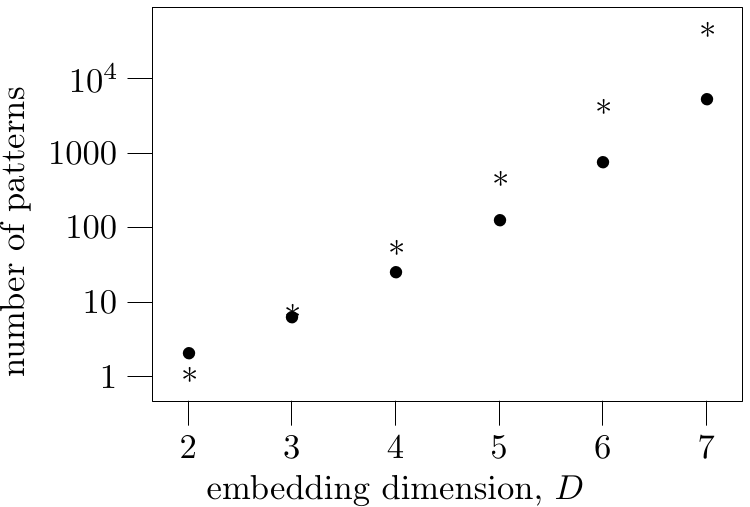}
\end{center}
\caption{\label{comp}Comparison of possible number of patterns. ({\small $\bullet$}) --- equal values are treated as described in Sec. \ref{equal0}, $D!$;  ($\ast$) ---  additional possible
 patterns due to  equal values, $N(D)$ graph, Eq. (\ref{ND}).}
\end{figure}

Any new pattern appears from embedding $D$-vector with $d$ different com\-po\-n\-ents, where
$d\in\{1,2,\dots,D-1\}$. So, having $d$ fixed, the number of corresponding new patterns is equal to the number
$N(D,d)$ 
of base-$D$ $D$-digit integers constructed from digits $\{0,1,\dots,d-1\}$ in such a way that
each of the $d$ digits is used at least once. This number can be calculated as
\newcommand{\Stirling}[2]{\genfrac{\{}{\}}{0pt}{}{#1}{#2}}
\newcommand{\stirling}[2]{\genfrac{\{}{\}}{0pt}{1}{#1}{#2}}
$$%\begin{equation}\label{Sd}
N(D,d)=d!\,\Stirling{D}{d},
$$%\end{equation}
where $\stirling{D}{d}$ --- is the Stirling numbers of the second kind, 
\cite[Part 5, \S 2]{Riordan1958}. Considering all possible values for $d$, we have for the 
total number of possible new patterns%
%\footnote{Compare this with \cite[Eq. (37)]{Haruna2013}.}%
:
\begin{equation}\label{ND}
N(D)=\sum\limits_{0<d<D}d!\,\Stirling{D}{d} = b(D) - D!\,,
\end{equation}
where $b(D)$ are known as ordered Bell numbers, see \cite[p.337]{Pippenger2010}
for naming discussion.
%%%maxima declare(D,integer); N(D):=sum(d!*stirling2(D,d),d,1,D-1); makelist(N(D),D,2,6);
%[1, 7, 51, 421, 3963]
%%%maxima makelist(D!,D,2,6,7);
%[2, 6, 24, 120, 720]
Calculating\footnote{The Stirling numbers were calculated with {\tt stirling2(D,d)}
function in the ``maxima'' computer algebra system ({\tt http://maxima.sourceforge.net/}).}
  $N(D)$ for $D\in\{2,3,4,5,6,7\}$ we see that the number of new patterns is 
normally greater than $D!$, see Fig. \ref{comp} and also Table \ref{T0}. Of course, the possible new patterns may only be
 significant when they can be observed (see discussion about this in \cite{Cuesta-Frau2018}).
 This depends on the process under study and embedding method.

\begin{table}
\begin{center}
\begin{tabular}{c c c c c c c}
\hline
$D$ & 2 & 3 & 4 & 5 & 6 & 7 \\
\hline
$N(D)+D!$ & 3 & 13 & 75 & 541 & 4683 & 47293 \\
\hline
\end{tabular}
\end{center}
\caption{\label{T0} Total number of possible patterns in the AE symbolization.}
\end{table}

\subsection{Coding}\label{cod}

Certainly, there are several possible implementations of the algorithm discussed
in Secs. \ref{Sym}, \ref{Ari}. Here, the one used for the examples in Sec. \ref{Exa},
and Appendix \ref{more}, 
below, is shown. It is a {\tt C++} program.
 It is expected that the sequence (\ref{raw}) is organized into a
one-dimensional array {\tt X[N]}. For calculating arithmetic order pattern
of vector $V_i$ shown in (\ref{Dvec}) it is necessary to pass
a pointer to the {\tt X[i]} to the function
{\tt get\_numerical\_pattern}, below, as its third argument:
{ \tt data\_point = X + i}. 

In the below example, {\tt X[i]} is declared as {\tt double}, but it can be of any 
type with appropriate sorting defined. 
The returning value is declared as {\tt mpz\_class}, which is a 
GNU multiple precision integer ({\tt https://gmplib.org/}). This is used because
for embedding dimensions $D>15$ the returned number representing an order pattern
may exceed 64 bits in size\footnote{It makes sense to use large embedding dimensions only for very long sequences of data. Otherwise, any observed pattern appears only once, which is unfavorable for estimating 
probabilities.}. For smaller $D$, {\tt mpz\_class} can be replaced with
{\tt int}, or {\tt long} everywhere in the code.

%\begin{lstlisting}
\begin{Verbatim}[numbers=left]
#include <gmp.h>
#include <gmpxx.h>
#include <forward_list>

/** 
    Function calculates numerical representation of order pattern
    of an embedding vector V_i = {x_i, x_{i + tau}, ...}. 
    Here D is the embedding dimension, tau is the delay. 
    The data_point points to the first component of the Vi in the 
    array of raw data.
*/
mpz_class get_numerical_pattern(int D,int tau,double * data_point)
{
 int k;
 std::forward_list<double> FL;
 auto it = FL.before_begin();
 for (k=0;k<D;k++) it = FL.emplace_after(it, data_point[k*tau]);
 FL.sort();
 FL.unique();
	
 int * pDpnm = new int [D]; // order pattern will be here
 int tag = 0;
 for (auto it = FL.begin(); it != FL.end(); ++it)
    {
     for (k=0;k<D;k++) 
        if (*it == data_point[k*tau]) pDpnm[k] = tag;
     tag++;
    }
	
 mpz_class pnum = 0; // arithmetic order pattern (initial value)
 mpz_class digval = 1; // initial value of a single digit
 for (k=0;k<D;k++)
    {
     pnum += pDpnm[k]*digval;
     digval *= D;
    }
 return pnum;
}
\end{Verbatim}
%\end{lstlisting}
This code is transparent and does not refer to combinatorics. 
At the same time, provided an embedding vector does not have equal components,
when loop at lines {\footnotesize 23-28} above is complete,
we obtain in the array \verb-pDpnm[D]- a permutation $\pi^{-1}$, where 
$\pi$ is the permutation for that vector, obtained in accordance with the
standard rules of \cite{Bandt2002} reproduced in Sec. \ref{P} with (\ref{asc}) adopted.

\section{Example}\label{Exa}
The discussed methodology has been tested at two surrogate sequences.
The purpose was to demonstrate that for a pair of sequences the standard permutation
entropy method gives roughly the same  entropy, whereas the arithmetic
entropy may be considerably different.

For calculating standard permutation entropy in situation when equal components in embedding
vectors are possible we replace the following fragment:

\begin{verbatim}
for (auto it = FL.begin(); it != FL.end(); ++it)
   {
    for (k=0;k<D;k++) 
       if (*it == data_point[k*tau]) pDpnm[k] = tag;
    tag++;
   }
\end{verbatim}
in the code of Sec. \ref{cod}, above, with the following one:\bigskip

\begin{verbatim}
for (auto it = FL.begin(); it != FL.end(); ++it)
   {
    for (k=D-1;k>=0;k--) 
       if (*it == data_point[k*tau]) pDpnm[k] = tag++;
   }
\end{verbatim}
With such a replacement we get in the array {\tt pDpnm[D]} 
above, the permutation, which is inverse to
one obtained for $V_i$ in the standard permutation entropy symbolization with
rules (\ref{asc}) and (\ref{rule}) adopted. 
As it was mentioned above, usage of inverse permutations instead of the initial ones
delivers the same value for the standard permutation entropy.

The two sequences, S1 and S2 are obtained as follows.
By means of function \verb-gsl_rng_uniform_int- from the GNU Scientific Library, 
\cite{Galassi2009},
we generate random numbers from the set $\{0,1,\dots,4\}$, which are equally probable.
Each obtained random number \verb-val- is written into the S1. The same number is
written into the S2, provided it is not equal to the number written to S2 at the
previous step. If it does, then the number (\verb-val- + 1) ({\bf mod} 5) is written
instead. This introduces a non-zero correlation between consecutive values in S2.
E.g., in the S2 any two consecutive values are always different. Examples of S1, S2
are as follows
$$
\text{S1} = (2,2,0,3,4,0,1,3,4,4,0,3,3,2,2,4,4,2,0,1,\dots),
$$$$
\text{S2} = (2,3,0,3,4,0,1,3,4,0,1,3,4,2,3,4,0,2,0,1,\dots).
$$
1\,000\,000 long S1 and S2 were produced and both permutation and arithmetic entropy
have been calculated. The results are shown in Tables \ref{T1} and \ref{T2}.

\begin{table}
\begin{center}
\begin{tabular}{c c c |c c c}
\hline
$D=3$ & PE & AE &$D=4$ & PE & AE \\
\hline
%S1    & 1.73092   & 2.55355   & S1   & 3.04323   &  4.27359   \\
S1    &  2.497  & 3.684   & S1   &  4.390  &  6.165   \\
\hline
%S2    & 1.64115   & 2.02343   & S2   & 2.90228   &  3.63053   \\
S2    &  2.368  & 2.919   & S2   &  4.187  &  5.238   \\
\hline
\end{tabular}
\end{center}
\caption{\label{T1}Comparison of permutation entropy (PE) and arithmetic entropy (AE) for embedding delay
$\tau=1$. Entropy is given in bits.}
\end{table}

\begin{table}
\begin{center}
\begin{tabular}{ccc|ccc}
\hline
$D=3$ & PE & AE &$D=4$ & PE & AE \\
\hline
%S1    &  1.73171  &  2.55333  & S1   & 3.04467   &  4.27364   \\
S1    &   2.498 &  3.684  & S1   & 4.393   &  6.166   \\
\hline
%S2    &  1.66819  &  2.54818  & S2   & 2.92817   &  4.22715   \\
S2    &  2.407  &  3.676  & S2   & 4.224   &  6.098   \\
\hline
\end{tabular}
\end{center}
\caption{\label{T2}Same as Table \ref{T1} for embedding delay $\tau=2$.}
\end{table}

Notice that arithmetic entropy is considerably greater than the permutation one.
This is due to high frequency of embedding vectors with equal components.
Also, from Table \ref{T1} with $\tau=1$ it can be seen that arithmetic entropy discriminates
 better between S1 and S2.
Although, case with delay $\tau=2$ shown in Table \ref{T2} is not similarly conclusive.
This might be due to construction method of the S2 sequence.
Namely, by pulling from S2 embedding vectors with delay 2, we may get vectors with equal adjacent
components, similarly to S1 case. This alleviates difference between S1 and S2.
For $\tau=1$, embedding vectors for S2 do not have equal adjacent components. More
examples are in the Appendix \ref{more}, below.

\section{Conclusions and discussion}
In this note, we have discussed a method for calculating entropy in a sequence of data,
which is similar to the permutation entropy method. The characteristic features of this 
method are as follows: 
\begin{itemize}
\item[(i)] it treats equal components in the embedding vectors as being
equal instead of ordering them artificially;
\item[(ii)] it is entirely exempt of combinatorics, labeling order patterns by integers instead of permutations;
\item[(iii)] if embedding vectors do not have equal components, this method delivers exactly the same
value for the entropy as does the standard permutation entropy one.
\end{itemize}

In the symbolization procedure discussed in Sec. \ref{Sym}, new order patterns may
appear as compared to the standard permutation method, see Sec. \ref{got}, above. 
Those new patterns arise from embedding vectors with some components being equal to each other.
In the standard permutation entropy method, 
the embedding vectors characterized by those new patterns, if any, 
are labeled by permutations as if there were no equal components. 
This is made possible through ordering equal values in accordance with the rule (\ref{rule}). 

Mathematically, replacing embedding vectors with their
 order patterns means constructing a quotient set from
the set of all embedding vectors with respect to some equivalence relation,
\cite{Keller2007,Bian2012,Piek2019}. In the case of permutation entropy, the corresponding equivalence relation
is defined by (\ref{rule}) and either (\ref{desc}), or (\ref{asc}).
Denote it by $\sim_P$. 
For arithmetic entropy, the corresponding equivalence relation
is defined by the algorithm described in the first paragraph of Sec. \ref{Sym}.
Denote it by $\sim_A$. It is clear that for two embedding vectors $U$, $V$,
if
$U\, \sim_A\,V$, then $ U\, \sim_P\,V$. 
Namely, if $U$, $V$ have the same arithmetic order pattern
then they do have the same permutation order pattern.
That means that $\sim_P$ is coarser relation than $\sim_A$. 
Other equivalence relations could be offered, which are courser than $\sim_P$,
or finer than $\sim_A$, or lying in between, or incomparable with the both, 
see e.g. \cite{Berger2017}.
A symbolization which still uses permutations, but is equivalent to discussed here,
as regards the treatment of equal values in embedding vectors,
has been proposed in \cite{Haruna2013}, see discussion in the Appendix B, below.
Which one is better depends on the data sequence and which kind of redundancy
is intended to strip.\bigskip\medskip

{
\small
\appendix
\section{Equivalence with permutations}\label{pi-1}
The following theorem proves the statement made in Sec. \ref{Sym}.
\begin{theorem}
Suppose, that an embedding vector $V=(x_0,x_1,\dots,x_{D-1})$ does not have equal components.
Then its symbolic pattern $\alpha(V)$, obtained as described in Sec. \ref{Sym} after  Eq. (\ref{aldef}), above,
represents permutation which is inverse to the $\pi(V)$ --- the permutation obtained in the standard
permutation entropy approach with convention (\ref{asc}) adopted:
$$%\begin{equation}\label{api-1}
\alpha = \pi^{-1}.
$$%\end{equation}
\end{theorem}
\noindent{\bf Proof.} Since $V$ has no equal components then
$\alpha(V)=(a_0,a_1,\dots,a_{D-1})$ represents some permutation of sequence $(0,1,\dots,D-1)$. Further, the procedure of obtaining $\alpha(V)$
from $V$ does not change rank order: for any $0\le i,j<D$, if $x_i<x_j$ then $a_i<a_j$
and vice verse.
If so, then $\alpha(V)$ can be used for calculating standard permutation $\pi(V)$:
$$%\begin{equation}\label{equa}
\pi(V)=\pi(\alpha(V)).
$$%\end{equation}
In this course, after arranging elements of $\alpha$ as required in (\ref{asc})  one obtains:
\begin{equation}\label{arra}
(a_0,a_1,\dots,a_{D-1})\quad\to\quad (a_{r_0},a_{r_1},\dots,a_{r_{D-1}})
=
(0,1,\dots,D-1).
\end{equation}
Obtained permutation  $\pi = (r_0,r_1,\dots,r_{D-1})$ acts as follows:
$$i\to r_i,\quad i=0,1,\dots,D-1.$$
Now, take into account that $\alpha(V)$ has number $i$ at position $r_i$, see (\ref{arra}).
That means that order pattern $\alpha(V)$, if treated as a permutation, acts as follows:
$$i\leftarrow r_i,\quad i=0,1,\dots,D-1\,.$$
The latter just means that $\alpha(V)=(\pi(V))^{-1}$.

Due to this theorem, method discussed in this note is equivalent to the standard permutation 
entropy method if in any embedding vector any two components are different.

\section{Comparison with modified permutation entropy}\label{CMP}
Several versions of modified permutation entropy symbolization have been proposed.
We analyze here those proposed in \cite{Bian2012} and \cite{Haruna2013}.
Consider firstly \cite{Bian2012}. The symbolization proposed there is obtained as follows.
Having an embedding vector $V=(x_0,x_1,\dots,x_{D-1})$ arrange its components as shown in 
(\ref{asc}), above, with their subscripts retained. 
If there are equal components, arrange their subscripts similar to the (\ref{rule}) rule,
or any other way. 
Before fetching the row of subscripts 
in the resulting vector  $V_\pi=(x_{r_0},x_{r_1},\dots,x_{r_{D-1}})$ 
as modified symbol, do the following preparation. If there is a group of equal components
$x_{s_0}=x_{s_1}=\dots=x_{s_l}$ in the $V_\pi$, then replace all subscripts in this group
by the smallest among $\{s_0,s_1,\dots,s_l\}$. Do this with all groups of equal components in
the $V_\pi$. Use the row of subscripts in the
such way modified $V_\pi$ as the modified symbol of $V$. This way modified symbolization
retains some information about equal components in the $V$. Let us denote this type of
symbolization as MPE and corresponding symbol as $\mu(V)$.

By comparing values presented in the Table \ref{T0}, above, with the data of \cite[TABLE I]{Bian2012}
we see that the total number of possible patterns is bigger in the Table \ref{T0}. Therefore,
it could be expected that MPE symbolization used in \cite{Bian2012} is coarser than that discussed
in this note. Additional hint in the same direction is that for some embedding vectors
symbolization of \cite{Bian2012} gives the same result while method discussed in this note
gives two different. Here is one example: $V_1=(3,5,3,5)$, $V_2=(3,5,5,3)$.
MPE symbolization of \cite{Bian2012} gives 
both for $V_1$ and $V_2$ the same order pattern $(0,0,1,1)$, while AE symbolization gives
$(0,1,0,1)$ for $V_1$ and $(0,1,1,0)$ for $V_2$ resulting in two different numerical patterns,
68 and 20 calculated for $D=4$ as shown in (\ref{A}), above.
Notice now that for any two embedding vectors $V_1$ and $V_2$,
\begin{equation}\label{claim}
\text{\bf if } \alpha(V_1)=\alpha(V_2), \text{ \bf then }\mu(V_1)=\mu(V_2). 
\end{equation}
Indeed, MPE symbol of any vector $V$ is obtained through rearranging components of $V$ in accordance
to their rank order. Symbol $\alpha(V)$, if considered as a vector,  has the same rank of its components as does $V$.
Therefore, $\alpha(V)$ can be used for calculating $\mu(V)$ instead of $V$ itself.
If so, then (\ref{claim}) becomes evident.
The above reasoning proves that MPE and AE methods of symbolization are comparable and AE is finer than
MPE.\medskip

Consider now symbolization used for modified permutation entropy proposed in \cite{Haruna2013}.
In this symbolization each embedding vector $V$ is symbolized with $\sigma(V)$,
The $\sigma(V)$ has the following structure:
\begin{equation}\label{MPE2}
\sigma(V)= (\pi(V), \mathbf{e}(V)).
\end{equation}
Here, $\pi(V)$ is a permutation.
The second half in (\ref{MPE2}), $\mathbf{e}(V)=(e_1,e_2,\dots,e_{D-1})$, 
keeps information about equal components in $V$. 
Call this symbolization MPE2. The symbol $\sigma(V)$ is obtained as follows.
 Arrange components in $V=(x_0,x_1,\dots,x_{D-1})$ 
in the ascending order keeping their subscripts. As a result we obtain a sequence of groups
consisting of equal components. Each group may have from one to $D$ elements. 
Of course, in the latter case there will be only one group. The value composing each group
in the sequence
increases from left to right. Arrange subscripts in each group in the ascending order.
Denote this way prepared sequence of components with their initial subscripts as 
$\tilde{V}=(v_{r_0},v_{r_1},\dots,v_{r_{D-1}})$.
The row of subscripts in $\tilde{V}$ is $\pi(V)$ from (\ref{MPE2}). 
This is standard PE symbol with (\ref{asc}) adopted with the only difference that in
the rule (\ref{rule}) the opposite inequality sign is used.
The sequence $\mathbf{e}(V)$ is composed of $D-1$ zeros and ones by the following rule:
{\bf if} $v_{r_{i-1}}= v_{r_{i}}$ {\bf then} $e_i=1$, {\bf otherwise} $e_i=0$.
\begin{theorem}
Symbolization MPE2 produces the same partition of a set of embedding vectors
 as does the AE one described in the Sec. \ref{Sym}, above.
\end{theorem}
\noindent{\bf Proof.}
In order to prove this statement we need to show that for any $V_1$, $V_2$
the following equivalence holds:
\begin{equation}\label{alsig}
\alpha(V_1)=\alpha(V_2)\quad \Leftrightarrow\quad \sigma(V_1)=\sigma(V_2).
\end{equation}
It is easily seen that $\sigma(V)$ can be unambiguously
recovered from $\alpha(V)$. Indeed, $V$ and $\alpha(V)$ considered as a vector, 
 have the same rank order of 
components. And calculation of  $\sigma(V)$ is based exclusively on the rank order.
Therefore, 
\begin{equation}\label{sa}
\sigma(V)=\sigma(\alpha(V)).
\end{equation}
Thus, vectors with same $\alpha$ will have same $\sigma$. This proves one half of (\ref{alsig}).
In order to prove the second half, we need to show how $\alpha(V)=(a_0,a_1,\dots,a_{D-1})$ 
can be unambiguously
recovered from $\sigma(V)$. For this purpose we use the mentioned above equality (\ref{sa}).
So, if we arrange the $\alpha(V)$ in ascending order retaining subscripts, we obtain,
instead of $\tilde{V}$, above, a vector 
$\tilde\alpha=(\tilde a_{r_0},\tilde a_{r_1},\dots,\tilde a_{r_{D-1}})$.
This vector consists of groups of equal values: the first group has only zeros, the second
one --- only '1', the last one --- only '$d-1$', where $d$ is the number of unique components in
the $V$ or $\alpha(V)$. The sequence of subscripts in the $\tilde\alpha$ is the permutation,
which constitutes the first part in the $\sigma(V)$. If one would have an $\tilde\alpha$
without subscripts inherited from the $\alpha(V)$, 
in the form of $\beta=(b_0,b_1,\dots,b_{D-1})$, the required
$\alpha$ might be obtained by applying permutation $\pi(V)$ to $\beta$. Namely,
for $i=0,1,\dots,D-1$, $a_{r_i}=b_i$, where $r_i$ is taken from the permutation $\pi(V)$:
$r_i=\pi_i$.
The required  sequence $\beta$ can be recovered from the second
part of $\sigma$. For this purpose, do the following reprocessing of $\mathbf{e}(V)$. 
Replace $\mathbf{e}(V)$ with the following sequence $(0,e_1,e_2,\dots,e_{D-1})$.
With the obtained new sequence proceed as follows.
At the step number one, if $e_1=1$ replace it with 0, otherwise replace it wit 1. 
Similarly, at the step number $i$, if $e_i=1$ replace it with the number put at the previous step
in place of $e_{i-1}$. Otherwise, replace $e_i$ with that same number incremented by 1.
After replacing $e_{D-1}$ we obtain the required sequence $\beta$. This completes the proof.

\section{One more example\protect\footnote{See also  \cite[Sec. 4.1]{Zunino2017}}}\label{more}

Here we consider the sequence of digits in decimal expansion of $\sqrt{2}$.
The first 1 million digits in the decimal expansion of $\sqrt{2}$ has been downloaded from
here: \\
{\tt https://catonmat.net/tools/generate-sqrt2-digits} and here:\\
{\tt https://apod.nasa.gov/htmltest/gifcity/sqrt2.1mil}. Denote this sequence S1.
The first 10 millions digits in the decimal expansion of $\sqrt{2}$ has been downloaded from
here: {\tt https://apod.nasa.gov/htmltest/gifcity/sqrt2.10mil}. Denote this sequence S10.
Both PE and AE were calculated for both S1 and S10 for different embedding dimensions $D$
with delay $\tau=1$. The $D$ values were chosen based on the number of occurrences
of different order patterns in the S1, see Table \ref{occur}. Based on the data of Table \ref{occur}
we skip $D=6$ and $D=7$ cases because the number of occurrences of some 
arithmetic entropy patterns is too small
for calculating probabilities. The numbers obtained for entropy  are presented in Tables \ref{2S1} and \ref{2S10}.
\begin{table}
\begin{center}
\begin{tabular}{c c c c c c c}
\hline
D         & 2 & 3 & 4 &5 & 6 &7 \\
\hline
$n_{min}$   &  100169   & 10055   & 1024     &  105  &  14   &  1\\
\hline
$n_{max}$    &  450143  & 120545   & 21274   &  2661  &  300  &   47\\
\hline
$N_{tot}$     &  3      & 13       & 75      &  541  &   4683  &  47293\\
\hline
\end{tabular}
\end{center}
\caption{\label{occur}Occurrences of different AE patterns in S1. Here $n_{min}$ denote the smallest
number of repetitions in S1 for a pattern, $n_{max}$ --- is the biggest number of repetitions, 
$N_{tot}$ --- is the total number of patterns found. Compare with Table \ref{T0}.}
\end{table}

\begin{table}
\begin{center}
\begin{tabular}{c c c c c }
\hline
D         &     2 &     3 &     4 &    5\\
\hline
AE       &   1.369  & 3.477   &  6.067    &  8.992 \\
\hline
PE       &  0.993  &  2.563 &  4.536  &   6.816 \\
\hline
NAE         &  0.864     &   0.940     &   0.974   &  0.990 \\
\hline
NPE         &  0.993     &   0.992     &   0.989   &  0.987 \\
\hline
\end{tabular}
\end{center}
\caption{\label{2S1}Arithmetic entropy, permutation entropy and normalized entropies for the first
1\,000\,000 digits of $\sqrt{2}$. Entropy is given in bits. NAE is calculated as AE/$\log_2(N_{tot})$, 
were $N_{tot}$ is taken from the bottom row of Table \ref{occur}.}
\end{table}

\begin{table}
\begin{center}
\begin{tabular}{c c c c c }
\hline
D         &     2 &     3 &     4 &    5\\
\hline
AE       &   1.369  & 3.476   &  6.066    &  8.991 \\
\hline
PE       &  0.993  &  2.563 &  4.537  &   6.817 \\
\hline
NAE         &    0.864     &   0.939     &  0.974    &  0.990 \\
\hline
NPE         &  0.993     &   0.992     &   0.990   &  0.987 \\
\hline
\end{tabular}
\end{center}
\caption{\label{2S10}Same as in Table \ref{2S1} for first 10\,000\,000 digits.}
\end{table}
This data, which are obtained numerically, can be checked analytically.
Indeed, the number $\sqrt{2}$ is believed to be base 10 normal \cite{Queffeec2006}. This means 
that any combination of $n$ digits can be found in the expansion with probability $10^{-n}$.
For example, if $n=2$, there are 10 combinations $\{(0,0),(1,1),\dots,(9,9)\}$ with AE pattern 
$\alpha_0=(0,0)$, 45 combinations $\{(0,1),(0,2),\dots,(0,9),(1,2),\dots,(1,9),(2,3),\dots,(8,9)\}$ 
with AE pattern $\alpha_1=(0,1)$, and the same amount with AE pattern $\alpha_2=(1,0)$.
This gives for the probabilities: $p(\alpha_0)=0.1$, $p(\alpha_1)=0.45$, $p(\alpha_2)=0.45$.
And $\text{ AE}_2=-0.1\log_2(0.1)-2\cdot 0.45\log_2(0.45)= 1.369$.
%maxima (-0.1*log(0.1)-2*0.45*log(0.45))/log(2),numer;
In the PE symbolization, both $\alpha_0$ and $\alpha_1$ correspond to $\pi_1^{-1}=(0,1)$
and $\alpha_2$ corresponds to $\pi_2^{-1}=(1,0)$
(we use here the rule (\ref{rule}) with inverse inequality sign).
 This gives 
$\text{ PE}_2=-0.55\log_2(0.55)- 0.45\log_2(0.45)= 0.993$.
%maxima (-0.55*log(0.55)-0.45*log(0.45))/log(2),numer;

From the Tables \ref{2S1} and \ref{2S10} we see that AE is usually bigger than PE.
This could be explained by the bigger total number of patterns available in the AE symbolization.
Perhaps, for this same reason normalized AE is smaller than NPE for small $D$.
What seems unexpected, it is the opposite behavior of NPE and APE with growing $D$.
Namely, NAE is increasing and NPE is decreasing function of $D$ for the parameter set 
considered\footnote{$D=7$ and 8 were considered for S10. The results, as regards decreasing
and increasing, support those observed for smaller $D$.}.
As it is illustrated in the previous paragraph, the $D$-tuples of digits from the expansion
sequence are distributed unevenly between different order patterns both for PE and AE. 
(This might explain dispersion of the patterns' frequencies observed in \cite[Fig. 8]{Zunino2017}).
The above mentioned behavior with increasing $D$ suggests that the unevenness decreases
for AE and increases for PE, at least in some ``normalized'' sense. 
This is for the $\sqrt{2}$ expansion. Whether a similar behavior takes place for other sequences,
and a possible practical utilization
of this fact require additional study.

}\bigskip

\noindent{\large\bf Data Availability}\smallskip

\noindent
{\small The data used to support the findings of this study are available from the author upon request.}
\bigskip

\noindent{\large\bf Conflicts of Interest}\smallskip

\noindent
{\small The author declares that there are no conflicts of interest.}
\bigskip

\noindent{\large\bf Acknowledgments}\smallskip

\noindent
{\small In this note the following free software have been used:
(i) linux operating system ({\tt https://getfedora.org/}); (ii) GNU Scientific Library,
\cite{Galassi2009},\\
({\tt https://www.gnu.org/software/gsl/}); (iii) GNU Multiple Precision Arithmetic Library
({\tt https://gmplib.org/}); (iv) Maxima, a free Computer Algebra System\\ ({\tt http://maxima.sourceforge.net/}); (v) RefDB, a free Reference Manager 
created by Markus Hoenicka
({\tt http://refdb.sourceforge.net/}).
The present work was partially supported by the Program of Fundamental
Research of the Department of Physics and Astronomy of the National Academy
of Sciences of Ukraine "Mathematical models of nonequilibrium processes in
open systems" N 0120U100857.
}

%\bibliographystyle{unsrt} 

% getref -t bibtex -o /home/alex/texts/texts.my/journals.my/Complexity/2rev4/References.bib :CK:=Bandt2002 OR :CK:=Gutjahr2020 OR :CK:=Zanin2012 OR :CK:=Piek2019 OR :CK:=Keller2014 OR :CK:=Bandt2005 OR :CK:=Keller2007  OR :CK:=Chen2019 OR :CK:=Porta2007 OR :CK:=Aziz2005 OR :CK:=Zunino2017 OR :CK:=Porta2001 OR :CK:=Bian2012 OR :CK:=Berger2017 OR :CK:=Bandt2019 OR :CK:=Bandt2017 OR :CK:=Cuesta-Frau2018 OR :CK:=Tylov2018 OR :CK:=Pippenger2010 OR :CK:=Riordan1958 OR :CK:=Bariviera2015 OR :CK:=Azami2016 OR :CK:=Galassi2009 OR :CK:=Haruna2013 OR :CK:=Kulp2014 OR :CK:=Queffeec2006 OR :CK:=

%\bibliography{References}{}

\end{document}